\documentclass[11pt]{article}
\usepackage{amsmath,amssymb,color,graphics,epsfig}

\usepackage{amsfonts}

\newcommand{\hoch}[1]{$\, ^{#1}$}

\newcommand{\be}{\begin{equation}}
\newcommand{\ee}{\end{equation}}
\newcommand{\bea}{\setlength\arraycolsep{2pt} \begin{eqnarray}}
\newcommand{\eea}{\end{eqnarray}}

\def\ft#1#2{{\textstyle{\frac{\scriptstyle #1}{\scriptstyle #2} } }}
\def\fft#1#2{{\frac{#1}{#2}}}

\def\0{{\sst{(0)}}}
\def\1{{\sst{(1)}}}
\def\2{{\sst{(2)}}}
\def\3{{\sst{(3)}}}
\def\4{{\sst{(4)}}}
\def\5{{\sst{(5)}}}
\def\6{{\sst{(6)}}}
\def\7{{\sst{(7)}}}
\def\8{{\sst{(8)}}}
\def\sst#1{{\scriptscriptstyle #1}}
\def\oneone{\rlap 1\mkern4mu{\rm l}}

\begin{document}


\begin{center}
{\large {\bf Lifshitz and Schr\"odinger Vacua, Superstar Resolution \\
in Gauged Maximal Supergravities}}

\vspace{10pt}
Hai-Shan Liu\hoch{1} and H. L\"u\hoch{2}

\vspace{10pt}

\hoch{1} {\it Institute for Advanced Physics \& Mathematics,\\
Zhejiang University of Technology, Hangzhou 310023, China}

\vspace{10pt}

\hoch{2}{\it Department of Physics, Beijing Normal University,
Beijing 100875, China}

\vspace{40pt}

\underline{ABSTRACT}
\end{center}

We consider the subset of gauged maximal supergravities that consists of the $SO(n+1)$ gauge fields $A^{ij}$ and the scalar deformation $T^{ij}$ of the $S^n$ in the spherical reduction of M-theory or type IIB.  We focus on the Abelian Cartan subgroup and the diagonal entries of $T^{ij}$.  The resulting theories can be viewed as the STU models with additional hyperscalars.  We find that the theories with only one or two such vectors can be generalized naturally to arbitrary dimensions.  The same is true for the $D=4$ or 5 Einstein-Maxwell theory with such a hyperscalar. The gauge fields become massive, determined by stationary points of the hyperscalars {\it a la} the analogous Abelian Higgs mechanism. We obtain classes of Lifshitz and Schr\"odinger vacua in these theories.  The scaling exponent $z$ turns out to be rather restricted, taking fractional or irrational numbers.  Tweaking the theories by relaxing the mass parameter or making a small change of the superpotential, we find that solutions with $z=2$ can emerge.  In a different application, we find that the resolution of superstar singularity in the STU models by using bubbling-AdS solitons can be generalized to arbitrary dimensions in our theories.  In particular, we obtain the smooth AdS solitons that can be viewed as the resolution of the Reissner-Nordstr\o m superstars in general dimensions.

\vfill {\footnotesize Emails: hsliu.zju@gmail.com\ \ \ mrhonglu@gmail.com}

\thispagestyle{empty}

\pagebreak

\tableofcontents
\addtocontents{toc}{\protect\setcounter{tocdepth}{2}}



\section{Introduction}

Lifshitz and Sch\"odinger spacetimes \cite{Kachru:2008yh,Son:2008ye, Balasubramanian:2008dm} are primary candidate backgrounds for studying non-relativistic condensed matter systems via the AdS/CFT correspondence.  These systems exhibit a dynamical Lifshitz scaling near some critical fixed points
\begin{equation}
t\rightarrow \lambda^z t\,,\qquad x_i\rightarrow \lambda x_i\,,\qquad
z\ne 1\,.
\end{equation}
Instead of the conformal invariance corresponding to having $z=1$, the temporal and spatial coordinates scale anisotropically.
Lifshitz and Sch\"odinger geometries embed such anisotropic scaling
invariance in one dimension higher via radial foliation with the radial coordinate $r$ scaling as $r\rightarrow \lambda^{-1} r$. (In the case of Schr\"odinger geometry, there exists an additional internal null direction $\xi$ that scales as $\xi\rightarrow \lambda^{2-z} \xi$.) They are homogenous generalization of the anti-de Sitter spacetime (AdS) in planar coordinates, and reduce to the AdS when the scaling exponent $z$ becomes unit.

The theories proposed initially in \cite{Kachru:2008yh,Son:2008ye, Balasubramanian:2008dm} that admit the Lifshitz and Schr\"odinger vacua
do not have fundamental origins in string or M-theory, where the AdS/CFT correspondence is better established.  There has been great succuss in the embedding of these vacua in string and M-theory \cite{Hartnoll}-\cite{Mallayev:2013pwa}.  However, there is yet an example of such vacua in lower-dimensional (gauged) supergravities.  The embedding of the Lifshitz or Schr\"odinger solutions in $D=10$ string or $D=11$ M-theory typically involves additional transverse directions that cannot be dimensionally reduced to give a clean Lifshitz or Schr\"odinger solution in some lower-dimensional supergravity.

One goal of this paper is to construct a class of Lifshitz and Schr\"odinger solutions in gauged maximal supergravities. Gauged maximal supergravities can be obtained from the consistent Kaluza-Klein sphere reduction of eleven-dimensional supergravity or type IIB supergravity.
In particular, the $S^7$ \cite{de Wit:1986iy} and $S^4$ \cite{Nastase:1999cb,Nastase:1999kf} reductions of eleven-dimensional supergravity give rise to gauged maximal supergravities in $D=4$ and $D=7$ respectively; the $S^5$ reduction of type IIB supergravity is expected to give rise to gauged maximal supergravity in $D=5$, although the full reduction ansatz has not been given.  The explicit reduction ans\"atze for gauged half-maximal supergravities can be found in \cite{Lu:1999bc,Cvetic:1999un,Lu:1999bw,Cvetic:1999au}.  Also known is the full $S^5$ reduction ansatz for the $SL(2,R)$ singlet of type IIB supergravity \cite{Cvetic:2000nc}.  The consistency of sphere reduction appears to be closely related to the global symmetry enhancement associated with the torus reduction where the sphere is replaced by the torus of the same dimensions \cite{Cvetic:2000dm}.  The reduction ans\"atze can be complicated and theory dependent.  The fact that the full ansatz of the $S^5$ reduction of the type IIB theory is yet unknown demonstrates the point.  However, by examining various known reduction ansatz in the literature, we find that there is a universal description of the embedding for a subset of fields, namely the $SO(n+1)$ gauge fields $A^{ij}$ and all the scalar deformation $T^{ij}$ of the $S^n$.  The consistency of the reduction requires an additional constraint that is easily satisfied for the static charged solutions we construct in this paper.

We focus on the gauge potentials that are associated with the Cartan subgroup of the $SO(n+1)$ and the diagonal entries of $T^{ij}$, which is a further consistent truncation.  The theories describe STU models in gauged supergravities with additional hyperscalars $\varphi_i$ \cite{Chong:2004ce}.  The vector potentials acquire massive terms with the masses determined by the stationary points of the scalars $\varphi_i$, analogous to the Abelian Higgs mechanism.  In the AdS vacua, where $\varphi_i$'s vanish, the vector fields are massless,  However, we show that there exist other vacua where $\varphi_i$'s are non-vanishing, and hence the vector fields become massive.  This allows us to construct both Lifshitz and Schr\"odinger vacua in these theories, with some specific scaling exponent $z$.

In section 2, we give the ansatz for the $S^n$ reduction focusing only on the gauge fields $A^{ij}$ and the scalar deformation $T^{ij}$.  We obtain the lower-dimensional Lagrangian and the consistency constraint. Our result is a summary of the previously-known individual examples of consistent sphere reductions.  One can then perform a further consistent truncation to the Cartan subgroup with $A^1=A^{12}$, $A^2=A^{34}$, {\it etc.}~and the diagonal entries of $T^{ij}$.  The resulting theories are the STU models in gauged supergravities with additional hyperscalars $\varphi_i$.  Each vector $A_i$ acquires a massive term with mass $\sinh^2\varphi_i$.  This procedure of acquiring the mass for $A^i$ is analogous to the Abelian Higgs mechanism, but with more complicated scalar potentials. This truncation was obtained in \cite{Chong:2004ce}. The new result we obtain is that for the case with only $A^1$ and/or $A^2$ non-vanishing, we can generalize the Lagrangian in a natural way to arbitrary dimensions, with the essential properties kept.  In four or five dimensions, when all the $U(1)^4$ or $U(1)^3$ vectors of the STU modes are set to equal, we find that a hyperscalar $\varphi$ with a specific scalar superpotential can be introduced to the Einstein-Maxwell theory.  Furthermore, we obtain a natural higher-dimensional generalization of this system.

In section 3, we consider cosmological Einstein gravity with a massive vector and we review the construction of Lifshitz and Schr\"odinger solutions in this theory.  This provides the general relation how the exponent $z$ is related to the massive parameter and the cosmological constant.  In section 4, we construct a class of Lifshitz and Schr\"odinger solutions for the single-vector theory in general dimensions; they are solutions in gauged maximal supergravities in the relevant lower dimensions.  We find no further Sch\"odinger solutions if we turn on more vector fields.  The equations become rather complicated for the Lifshitz ansatz when multiple vector fields are turned on.  We obtain solutions when all the vectors are set to equal.

Gauged maximal supergravities have only one non-trivial parameter, the gauge coupling constant $g$. Furthermore the scalar potential associated with diagonal entries of the $T^{ij}$ has few extrema.  It follows that the scaling exponent $z$ in our solutions is highly restricted, with fractional or even irrational numbers.  In section 5, we tweak the supergravity Lagrangian by either modifying the superpotential or relaxing the coupling constant of the mass term.  We find that Lifshitz or Schr\"odinger solutions with the scaling exponent $z=2$ can emerge in these supergravity-inspired theories.

     In section 6, we turn to a different application of the Lagrangians
obtained in section 2. It is well-known that the STU models admit multi-charged AdS black holes in gauged supergravities.  In the BPS limit, these solutions become ``superstars'' \cite{Myers:2001aq} that have naked singularity.  These singularities can be resolved and the solutions become smooth BPS solitons using the bubbling AdS technique \cite{Lin:2004nb}.  Some of these bubbling AdS configurations in M-theory or type IIB become smooth solitons in lower-dimensional gauged supergravities \cite{Chong:2004ce}.  In this section we show that the generalized $D$-dimensional theories of single or two-vector or equal-vector theories admit the bubbling-AdS type of solitons.  In particular this allows us to resolve the naked singularity of the Reissner-Nordstr\o m AdS superstars in arbitrary dimensions.  We conclude the paper in section 7.

\section{Sphere Reduction and Maximal Gauged Supergravites}

\subsection{Sphere reduction of M-theory and type IIB}

Eleven-dimensional supergravity and type IIB supergravity admit maximally supersymmetric vacua of (AdS$_7\times S^4$, AdS$_4\times S^7$) and AdS$_5\times S^5$ respectively.  The corresponding consistent $S^n$-reductions give rise to gauged maximal supergravities with $SO(n+1)$ gauge group in lower dimensions, with the maximally supersymmetric AdS vacua. For our purpose, we shall concentrate on the subset of fields, namely the $SO(n+1)$ gauge fields $A^{ij}$ and all the scalar deformation $T_{ij}$ of the $S^n$.  By examining the known examples of the explicit reduction ans\"atze in the literature (\cite{de Wit:1986iy}-\cite{Cvetic:1999au} and \cite{Chong:2004ce,Cvetic:2000nc}), we find that for this subset of fields, the reduction ansatz has the same general structure:
\begin{eqnarray}
d\hat s_{D+n}^2 &=& \Delta^{\alpha} ds_D^2 + g^{-2} \Delta^{\beta} T_{ij} D\mu^i D\mu^j\,,\cr
\hat G_{\sst{(D)}} &=& - g U \epsilon_{\sst{(D)}} + g^{-1} (T_{ij} {*D} T_{jk}) \wedge (\mu^k D\mu^i)\cr
 &&- \ft{1}{2g^2} T^{-1}_{ik} T^{-1}_{j\ell}
{*F^{ij}}\wedge D\mu^k\wedge D\mu^\ell\,,
\end{eqnarray}
where $\epsilon_{\sst{(D)}}$ is the volume form of the metric $ds_D^2$ and
\begin{eqnarray}
\mu^i\mu^i&=&1\,,\qquad \Delta=T_{ij} \mu^i \mu^j\,,\qquad
U=2T_{ij}T_{jk} u^i u^k - \Delta T_{ii}\,,\cr
D\mu^i &=& d\mu^i + g A^{ij} \mu^j\,,\qquad
F^{ij}= dA^{ij} + g A^{ik} \wedge A^{kj}\,,\cr
DT_{ij} &\equiv& dT_{ij} + g A^{ik} T_{k j} + g A^{jk} T_{ik}\,,
\end{eqnarray}
The constants $\alpha$ and $\beta$ are given by
\begin{eqnarray}
(D,n)=(7,4):&& (\alpha,\beta)=(\ft13,-\ft23)\,,\cr
(D,n)=(4,7):&& (\alpha,\beta)=(\ft23,-\ft13)\,,\cr
(D,n)=(5,5):&& (\alpha,\beta)=(\ft12,-\ft12)\,.
\end{eqnarray}
Note that $i=1,2,\cdots,(n+1)$ and $A^{ij}$ is antisymmetric in $i,j$.  The scalars $T_{ij}$ are symmetric in $i,j$, but with $\det(T_{ij})=1$, giving a total of $\fft12n(n+3)$ degrees of freedom.  The resulting lower-dimensional Lagrangians have a universal structure for all $D=4,5$ and $7$:
\begin{eqnarray}
{\cal L}_D &=& R{*\oneone} - \ft14 T_{ij}^{-1} {*D} T_{jk}\wedge T^{-1}_{k\ell} DT_{\ell i} - \ft14 T_{ik}^{-1} T^{-1}_{j\ell} {*F^{ij}\wedge F^{k\ell}} - V {*\oneone}\,,\cr
V&=& \ft12 g^2 \left(2T_{ij} T_{ij} - (T_{ii})^2\right)\,.\label{genlag}
\end{eqnarray}
Since we deal with only a subset of the fields of the gauged maximal supergravities, the reduction is in general not consistent. The consistency requires an additional constraint on the $U(1)$ fields
\begin{equation}
F^{[ij}\wedge F^{k\ell]}=0\,.\label{ffcons}
\end{equation}
For all the static charged solutions we consider in this paper, this condition is satisfied. To further simplify the Lagrangian, we consider the diagonal truncation of the system. We set all the fields to zero, except for
\begin{eqnarray}
&&A^{12}=A^1\,,\qquad A^{34}=A^2\,,\qquad A^{56}=A^3\,,\qquad etc.\,,\cr
&&T_{ij}={\rm diag}(X_1 e^{-\varphi_1}\,, X_1 e^{\varphi_1}\,,
X_2 e^{-\varphi_2}\,, X_2 e^{\varphi_2}\,,\cdots)\,,\label{diagtrunc}
\end{eqnarray}
where $X_i$'s satisfy $\prod X_i = 1$.

In other words, we keep the gauge fields associated with the Cartan subgroup and diagonal entries of $T_{ij}$.  This diagonal truncation from (\ref{genlag}) is consistent.  Note that for $n=2k$, the last entry of the diagonal $T_{ij}$ in (\ref{diagtrunc}) is $X_0=(X_1X_2\cdots X_k)^{-k}$ without a corresponding $\varphi$ factor.

It turns out that when we keep only the $A^1$ and $A^2$ fields, the theory can be generalized naturally to arbitrary dimensions.  In what follows, we list the theories organized by the number of $U(1)$ fields.

\subsection{Single-vector theory}

It is consistent to keep only the $A^{12}=A$ and the first two entries of the diagonal $T_{ij}$ in (\ref{diagtrunc}).  We find that the theory can be naturally generalized to arbitrary dimensions. The Lagrangian is given by
\begin{equation}
e^{-1}{\cal L} = R - \ft12(\partial\phi)^2 - \ft12 (\partial\varphi)^2 -
\ft14 e^{a \phi} F^2 - 2 g^2 \sinh^2\varphi\, A^2 - V(\phi,\varphi)\,,\label{singleVlag}
\end{equation}
where
\begin{equation}
a=\sqrt{\fft{2(D-1)}{D-2}}\,,\qquad ab=-\fft{2(D-3)}{D-2}\,.\label{ab}
\end{equation}
Here we have defined a constant $b$ for later purpose.  (Note that in this paper we define $e=\sqrt{-g}$ that appears in $e^{-1}{\cal L}$ only, which should not be confused with the exponential symbol.  Also we use the parameter $g$ to denote the gauge coupling constant, and it should not be confused with the determinant of the metric that appears only in $\sqrt{-g}$.)  The scalar potential $V(\phi,\varphi)$ can be expressed in terms of superpotential $W$
\begin{eqnarray}
V &=& \left(\fft{\partial W}{\partial \phi}\right)^2 +
\left(\fft{\partial W}{\partial \varphi}\right)^2 - \fft{D-1}{2(D-2)} W^2\,,\cr
W &=& \sqrt{2}g\, \Big(e^{-\fft12a\phi} \cosh\varphi - \ft{a}{b} e^{-\fft12 b\phi}\Big)\,.
\end{eqnarray}
Explicitly, we have
\begin{equation}
V=2g^2 \left(e^{-a\phi} \sinh^2 \varphi - \ft{2(D-1)}{D-3} e^{-\fft12(a+b)\phi} \cosh\varphi - \ft{2(D-1)}{(D-3)^2} e^{-b\phi}\right)\,.\label{singleAV}
\end{equation}
Note that $(\phi,\varphi)=(0,0)$ are stationary points for both $W$ and $V$, with
\begin{equation}
W_0=\fft{2\sqrt2(D-2)}{D-3}g\,,\qquad V_0=-\fft{4(D-1)(D-2)}{(D-3)^2} g^2\,,\label{w0v0}
\end{equation}
giving rise to the AdS vacuum.
It is worth noting that setting the gauging parameter $g=0$ simply yields the Kaluza-Klein theory in $D$ dimensions.

    For $D=4,5,7$, the theory is a subset of gauged maximal supergravity
discussed earlier.  In fact, the $D=6$ case can also be embedded in six-dimensional gauged ${\cal N}=1$ supergravity with some appropriate matter multiplet. The Lagrangian in general dimensions with $\varphi=0$ were studied in \cite{Wu:2011zzh,Liu:2012jra}.

\subsection{Two-vector theory}

The generalization to arbitrary dimensions is possible if we keep $A^{12}=A_1$, $A^{34}=A_2$ and first four entries of the diagonal $T_{ij}$ in (\ref{diagtrunc}).  The Lagrangian is given by
\begin{eqnarray}
e^{-1}{\cal L} &=& R - \ft12(\partial\phi_1)^2- \ft12(\partial\phi_2)^2- \ft12 (\partial\varphi_1)^2 - \ft12(\partial\varphi_2)^2\cr
&& -
\ft14 X_1^{-2} F_1^2 -\ft14 X_2^{-2} F_2^2- 2 g^2 \sinh^2\varphi_1\, A_1^2
- 2 g^2 \sinh^2\varphi_2\, A_2^2 - V\,,
\end{eqnarray}
where
\begin{equation}
X_i = e^{-\ft12 \vec a_i\cdot \vec \phi}\,,\qquad i=1,2\,,
\end{equation}
with
\begin{equation}
\vec a_i \cdot \vec a_j = 4 \delta_{ij} - \fft{2(D-3)}{D-2}\,.\label{aiaj}
\end{equation}
To present the scalar potential $V$, let us define
\begin{equation}
X_0=e^{-\ft12 \vec a_0 \cdot \vec \phi}\,,\qquad \hbox{with}\qquad
\vec a_0\cdot\vec a_0=\fft{2(D-3)^2}{D-2}\,,\qquad \vec a_0\cdot \vec a_i =  - \fft{2(D-3)}{D-2}\,.
\end{equation}
In fact, $\vec a_0$ can be given in terms of $\vec a_i$ by:
\begin{equation}
\vec a_0=-\ft12(D-3)(\vec a_1 + \vec a_2)\,.
\end{equation}
In other words
\begin{equation}
X_0=(X_1 X_2)^{-\fft{D-3}2}\,.
\end{equation}
The scalar potential $V$ is again expressible in terms of a superpotential
\begin{eqnarray}
V &=& \sum_{i=1}^2\Big[\left(\fft{\partial W}{\partial \phi_i}\right)^2 +
\left(\fft{\partial W}{\partial \varphi_i}\right)^2\Big] - \fft{D-1}{2(D-2)} W^2\,,\cr
W &=& \sqrt{2}g\, \Big(X_1 \cosh\varphi_1 + X_2\cosh\varphi_2 + \ft2{D-3} X_0\Big)\,.
\end{eqnarray}
The $V$ and $W$ have both the stationary point $\phi_i=0=\varphi_i$, with the same (\ref{w0v0}).  Note that if we set $g=0$, the two vectors $A_1$ and $A_2$ can be viewed as the Kaluza-Klein and string winding vectors in the $S^1$ reduction of the bosonic string.  (The constraint (\ref{ffcons}), which now becomes $F_1\wedge F_2=0$, implies that we can set the 3-form field strength to zero.)

     The condition (\ref{aiaj}) can be solved up to an arbitrary
orthonormal rotation.  One way to solve it is to let
\begin{equation}
\vec a_1=(a,0)\,,\qquad \vec a_2=(b, \sqrt{\ft{8}{D-1}})\,,
\end{equation}
where $a,b$ are given in (\ref{ab}). This choice makes it explicit how to reduce the two-vector theory to the single-vector theory, by letting $\phi_2=0=\varphi_2$ and $A_2=0$.  An alternative choice is to let
\begin{equation}
\vec a_1=(\sqrt{\ft{2}{D-2}}, \sqrt{2})\,,\qquad
\vec a_2=(\sqrt{\ft2{D-2}}, -\sqrt2)\,.\label{aichoice2}
\end{equation}
This parametrization makes it easier to set two vectors equal, by letting $\phi_2=0$, $\varphi_1=\varphi_2\equiv\varphi/\sqrt2$ and $A_1=A_2\equiv A/\sqrt2$, giving
\begin{eqnarray}
e^{-1}{\cal L} &=& R - \ft12(\partial\phi)^2- \ft12 (\partial\varphi)^2 -
\ft14 e^{\tilde a\phi} F^2 - 2 g^2 \sinh^2(\fft{\varphi}{\sqrt2})\, A^2 - V\,,\label{twoequallag}
\end{eqnarray}
where
\begin{eqnarray}
W &=& 2\sqrt{2}g\, \Big(e^{-\fft12\tilde a\phi} \cosh\varphi - \ft{\tilde a}{\tilde b} e^{-\fft12 \tilde b\phi}\Big)\,,\cr
V &=&\left(\fft{\partial W}{\partial \phi}\right)^2 +
\left(\fft{\partial W}{\partial \varphi}\right)^2 - \fft{D-1}{2(D-2)} W^2\cr
&=&-4g^2\left( e^{-\tilde a\phi} + \ft{4}{D-3} e^{-\ft12(\tilde a + \tilde b)\phi} \cosh(\fft{\varphi}{\sqrt2}) -
\ft{D-5}{(D-3)^2} e^{-\tilde b \phi}\right)\,,\label{twoequalV}
\end{eqnarray}
with
\begin{equation}
\tilde a = \sqrt{\ft{2}{D-2}}\,,\qquad \tilde a \tilde b=-\ft{2(D-3)}{D-2}\,.
\end{equation}
For $D=4,5,6,7$, the Lagrangian can be embedded in gauged half-maximal supergravities.  The two-vector theory with $\varphi_i=0$ was considered in \cite{Chow:2011fh}.  The two-equal-charge Lagrangian is a special case of a more general class of theories considered in \cite{Lu:2013eoa}.

\subsection{Three and four-vector theories}

The three-vector and four-vector theories are the truncations of five-dimensional and four-dimensional gauged supergravities respectively.  The theories can be read off from (\ref{genlag}) with the constraint (\ref{diagtrunc}) and were obtained in \cite{Chong:2004ce}.  For completeness, we give the Lagrangian also.  In terms of form Language, the three-vector theory is given by \cite{Chong:2004ce}
\begin{eqnarray}
{\cal L}_5 &=& R{*\oneone} - \ft12 \sum_{i=1}^3 {*d\varphi_i\wedge d\varphi_i} - \ft12 \sum_{\alpha=1}^2 {*d\phi_\alpha\wedge d\phi_\alpha} - \ft12 \sum_{i=1}^3 X_i^{-2} {*F_i}\wedge F_i\cr
&& - 2g^2 \sum_{i=1}^3 \sinh^2\varphi_i\, {*A_i\wedge A_i} - V {*\oneone} +
F_1\wedge F_2\wedge A_3\,,\label{d5lag}
\end{eqnarray}
where $X_i$ are related to $\phi_\alpha$ as
\begin{equation}
X_1=e^{-\fft{1}{\sqrt6} \phi_1 - \fft{1}{\sqrt2} \phi_2}\,,\quad
X_2=e^{-\fft{1}{\sqrt6} \phi_1 + \fft{1}{\sqrt2} \phi_2}\,,\quad
X_3=e^{\fft{2}{\sqrt6} \phi_1}\,,
\end{equation}
We find that the scalar potential can be expressed in terms of a super potential:
\begin{equation}
V = \sum_{\alpha=1}^2\left(\fft{\partial W}{\partial \phi_\alpha}\right)^2 + \sum_{i=1}^3
\left(\fft{\partial W}{\partial \varphi_i}\right)^2 - \fft23 W^2\,,\qquad
W = \sqrt{2}g\, \sum_{i=1}^3 X_i \cosh\varphi_i \,.
\end{equation}
The Lagrangian is given in form language because of the $FFA$ term.  The Lagrangian is also the full bosonic sector of the gauged ${\cal N}=1$ supergravity with two vector multiplet.  The bosonic theory can be consistently truncated from the $SO(6)$ gauged maximal supergravity without needing the additional constraint (\ref{ffcons}).

    The four-vector theory in four-dimensions is given by
\cite{Chong:2004ce}
\begin{equation}
e^{-1} {\cal L}_4 = R - \ft12 \sum_{i=1}^4 (\partial\varphi_i)^2- \ft12 \sum_{\alpha=1}^3 (\partial\phi)^2 - \ft12 \sum_{i=1}^4 X_i^{-2} F_i^2 - 2g^2 \sum_{i=1}^4 \sinh^2\varphi_i\, A^2 - V \,,\label{d4lag}
\end{equation}
where
\begin{equation}
X_1=e^{\fft12(-\phi_1-\phi_2-\phi_3)}\,,\quad
X_2=e^{\fft12(-\phi_1+\phi_2+\phi_3)}\,,\quad
X_3=e^{\fft12(\phi_1-\phi_2+\phi_3)}\,,\quad
X_4=e^{\fft12(\phi_1+\phi_2-\phi_3)}\,.
\end{equation}
The scalar potential is given by
\begin{equation}
V = \sum_{\alpha=1}^3\left(\fft{\partial W}{\partial \phi_\alpha}\right)^2 + \sum_{i=1}^4
\left(\fft{\partial W}{\partial \varphi_i}\right)^2 - \fft34 W^2\,,\qquad W = \sqrt{2}g\, \sum_{i=1}^4 X_i \cosh\varphi_i \,.
\end{equation}

\subsection{Einstein-Maxwell theory with a hyperscalar}

We find that in both four and five dimensions discussed in the previous subsection,  there exists a further consistent truncation
\be
A_i=\fft{A}{\sqrt{N}}\,,\qquad \varphi_i = \fft{\varphi}{\sqrt{N}}\,,
\ee
where $N=4$ and 3 respectively, together with letting all $\phi_\alpha=0$. The resulting theory has a natural generalization to arbitrary dimensions:
\be
{\cal L}_D=\sqrt{-g}\Big( R - \ft12(\partial\varphi)^2 - \ft14 F^2 -2 g^2 \sinh^2(\fft{\varphi}{\sqrt{N}})A^2 + \fft{2Ng^2}{D-3} (D-2 + \cosh\fft{2\varphi}{\sqrt{N}})\Big)\,,\label{einmaxscalar}
\ee
where $F=dA$ and
\be
N=\fft{2(D-2)}{D-3}\,.\label{Nvalue}
\ee
The superpotential is given by
\be
W=\sqrt2 N g \cosh\fft{\varphi}{\sqrt{N}}\,.\label{superpot}
\ee
The scalar $\varphi$ can be consistently truncated out yielding the usual Einstein-Maxwell theory. Note that the constant $N$ is integer only in $D=4$ and 5, indicating the number of basic $U(1)$ building blocks, and the theories can be embedded in the respective gauged maximal supergravities. In higher dimensions, the theory cannot be embedded in string or M-theory and interestingly $N$ is no longer an integer.

\section{Massive Vector}

As we have seen in the previous section, analogous to the Abelian Higgs mechanism, the gauge field $A_i$ of the Cartan supgroup in maximal supergravities acquire a mass term with the effective mass $m_i$ determined by the modulus scalars $X^0_i$ and $\varphi_i^0$:
\begin{equation}
m_i = 2 g X^0_i \sinh\varphi^0_i\,.
\end{equation}
For the (supersymmetric) AdS vacua, we have $\varphi^0_i=0$, and hence $m_i=0$, as one would have expected.  When $\varphi_i$ acquires a non-vanishing expectation value $\varphi_i^0$, $m_i$ becomes non-zero.

Einstein gravity with a cosmological constant coupled to a massive vector allows one to construct both Lifshitz and Schr\"odinger vacua.  In this section, we review such construction.  Let us consider in $D$ dimensions
\begin{equation}
e^{-1} {\cal L}_D=R + (D-1)(D-2)g^2 - \ft14 F^2 - \ft12 m^2 A^2\,.\label{toylag}
\end{equation}
The equations of motion are
\begin{equation}
R_{\mu\nu} + (D-1) g^2 g_{\mu\nu} = \ft12 m^2 A_\mu A_\nu  +\ft12 (F_{\mu\nu}^2 - \ft1{2(D-2)} F^2 g_{\mu\nu})\,,\qquad \nabla_\mu F^{\mu\nu}=m^2 A^\nu\,.
\end{equation}

{\noindent \bf Lifshitz solution:}\ \ The ansatz for the Lifshitz solution is given by
\begin{equation}
ds^2=\ell^2\Big(-r^{2z} dt^2 + \fft{dr^2}{r^2} + r^2 dx^i dx^i\Big)\,,\qquad A=q r^z dt\,,\label{lifsans}
\end{equation}
Substituting the ansatz into the equations of motion, one finds that
\begin{eqnarray}
&&\ell^2=\fft{(D-2)z}{m^2}\,,\qquad
m^2=\fft{(D-1)(D-2)^2g^2 z}{z^2 + (D-3)z +(D-2)^2}\,,\cr
&&q^2 = \fft{2(z-1)(z^2+ (D-3)z+(D-2)^2)}{(D-1)(D-2)g^2 z}\,.\label{toysol1}
\end{eqnarray}
We can without loss of generality treat $g$ as fixed, then $z$ can be determined as a quadratic function in terms of parameter $m^2$.  For each $m$, there can be two solutions for $z$. Although $m^2$ can be negative for non-asymptotic flat spacetimes such as AdS, we consider only $m^2> 0$ since the Lagrangian (\ref{toylag}) is simply the toy model for the gauged supergravities discussed previously for which $m_i^2$ are all non-negative.  It follows that we must have $z\ge 0$.  Furthermore, the reality condition for $A$ and hence $q$ implies that solution (\ref{toysol1}) is valid only for
\begin{equation}
z\ge 1\,.
\end{equation}
This condition restricts the parameter range of $m$.  For an $m$ value whose $z$ lies in the range $0<z<1$, the solution can be analytically continued to a Lifshitz-like anisotropic vacuum
\begin{eqnarray}
ds^2&=&\ell^2\Big(r^{2z} dy^2 + \fft{dr^2}{r^2} + r^2 dx^\mu dx_\mu\Big)\,,\qquad A=q r^z dt\,,\cr
q^2 &=& \fft{2(1-z)(z^2+ (D-3)z+(D-2)^2)}{(D-1)(D-2)g^2 z}\,,\qquad 0<z<1\,.\label{toyzle1}
\end{eqnarray}
with $m$ and $\ell$ still given in (\ref{toysol1}).  Now the anisotropic scaling occurs on one spatial direction rather than on the earlier temporal direction.  We shall see that this type of solutions also emerge in gauged maximal supergravities.

{\noindent \bf Schr\"odinger solution:}\ \ The ansatz is given by
\begin{equation}
ds^2=\ell^2\Big(-r^{2z} dt^2 + \fft{dr^2}{r^2} + r^2(-2dx dt + dy^i dy^i)\Big)\,,\qquad A=q r^z dt\,.\label{schrans1}
\end{equation}
The solution is
\begin{equation}
\ell=g^{-1}\,,\qquad m^2 = z(z+D-3) g^2\,,\qquad q^2=\fft{2(z-1)}{z g^2}\,.\label{toysol2}
\end{equation}
The reality condition requires that we must have $z\ge 1$ and $z\le -(D-3)$.  For $0<z<1$, the solution can be analytically continued to a real solution
\begin{eqnarray}
ds^2&=&\ell^2\Big(r^{2z} dt^2 + \fft{dr^2}{r^2} + r^2(-2dx dt + dy^i dy^i)\Big)\,,\qquad A=q r^z dt\,,\cr
q^2&=&\fft{2(1-z)}{z g^2}\,.
\end{eqnarray}
with $(\ell,m)$ the same as (\ref{toysol2}).  Both types of solutions can emerge in gauged maximal supergravities.

\section{Lifshitz and Schr\"odinger Vacua}

Having reviewed both the Lifshitz and Schr\"odinger vacua in Einstein gravity with a negative cosmological constant coupled to a massive vector field, we are now in the position to construct such solutions in gauged maximal supergravities obtained in section 2.

\subsection{Lifshitz vacua}

Let us first consider the single-vector theory given in section 2.2.  The ansatz for the Lifshitz solution is still given in (\ref{lifsans}) while the scalars $\phi$ and $\varphi$ are treated as constants.  Make a vielbein choice $e^{\bar 0}=r^z dt/\ell$, $e^{\bar r}=dr/(r\ell)$ and $e^{\bar i} = r dx^i/\ell$, we have
\begin{eqnarray}
&&R_{\bar 0\bar 0} =  z(z+ D-2)\ell^{-2}\,,\qquad
R_{\bar r\bar r} = -(z^2 + D-2) \ell^{-2}\,,\qquad
R_{\bar i\bar i} = - (z + D-2)\ell^{-2}\,,\cr
&&A^2=-q^2\ell^{-2}\,,\quad F_{\bar 0\bar 0}^2 = (q z)^2\ell^{-4}\,,\quad
F_{\bar r\bar r}^2=-(q z)^2\ell^{-4}\,,\quad F^2=-2 (qz)^2\ell^{-4}\,.
\end{eqnarray}
Here we use barred indices to denote tangent flat indices. The equation for $A$ implies that
\begin{equation}
z(D-2) e^{a\phi}= 4 g^2 \ell^2 \sinh^2\varphi\,.\label{singleom1}
\end{equation}
To obtain the equations for the scalars, we first define
\begin{eqnarray}
\tilde V &=& -
\ft14 e^{a\phi} F^2 - 2g^2 \sinh^2\varphi\, A^2 - V\cr
&=& \ft12 q^2 z^2 \ell^{-4} e^{a\phi} + 2 g^2 q^2 \ell^{-2} \sinh^2\varphi - V\,.
\end{eqnarray}
The scalar equations are then given by
\begin{equation}
\fft{\partial \tilde V}{\partial \phi} = 0 = \fft{\partial \tilde V}{\partial \varphi}\,.
\end{equation}
The Einstein equations of motion are
\begin{eqnarray}
z(z+D-2) &=& 2 g^2 q^2 \sinh^2\varphi - \ft{\ell^2 V}{D-2} +
\ft{(D-3)q^2 z^2}{2(D-2)\ell^2} e^{a\phi}\,,\cr
z^2+D-2 &=& - \ft{\ell^2 V}{D-2} +
\ft{(D-3)q^2 z^2}{2(D-2)\ell^2} e^{a\phi}\,,\cr
z+D-2 &=& - \ft{\ell^2 V}{D-2} -
\ft{q^2 z^2}{2(D-2)\ell^2} e^{a\phi}\,,\label{singleom2}
\end{eqnarray}
The system has a total of five parameters $(\phi,\varphi,\ell,q,z)$, but six equations, and hence {\it a priori} there may not be any solutions.  However, we find that a real solution exists in each given dimensions $D$:
\begin{eqnarray}
D=4:&& L=\{\sqrt3-1,\ft14(\sqrt3-1), \ft12(\sqrt3-2), 1, \sqrt3\}\,,\cr
D=5:&& L=\{\ft14(\sqrt{57}-3), (\ft{9}{256}(\sqrt{57}+21))^{\fft13}, \ft34, \sqrt{\ft13(\sqrt{57}-7)}, \sqrt{\ft12(\sqrt{57}-5)}\}\,,\cr
D=6:&& L=\{\ft43, 2^{\fft98} {11}^{\fft38} 3^{-\fft34}, \ft{44}{9}, \ft{3}{2\sqrt{22}}, \ft{5}{\sqrt{22}}\}\,,\cr
D=7:&&L=\{\sqrt6-1, (629+291\sqrt6)^{\fft15}, 2 (4 + \sqrt6),
\sqrt{\ft15(3\sqrt6-7)},\cr
&&\qquad\qquad\qquad \ft12\sqrt{3(\sqrt6-1)}\}\,,
\end{eqnarray}
where $L$ denotes the list of the following quantities
$L=\{z, g^2 \ell^2, g^2 q^2, e^{\fft2{a}\phi}, \cosh\varphi\}$.  Here we have presented the details only for $D=4,5,6,7$ associated with gauged supergravities. For general $D$, the solutions become too complicated to present.  Instead we give only the solution for $z$:
\begin{equation}
z=\ft{1}{4(D-3)}\Big(\sqrt{D^4+2D^3 - 47D^2+112 D - 32} -(D^2-7D+16)\Big)\,.
\end{equation}
Note that for $D=4$, we have $z<1$ and consequently, as discussed in section 3, we have $q^2<0$.  The solution can be real by an analytical continuation to the type (\ref{toyzle1}). For $D\ge 5$, we have $1<z<2$, and $z\rightarrow 2$ as $D\rightarrow \infty$.  The values for $z$ are all irrational except for $z=\fft43$ in $D=6$.

For the two vector theory, the ansatz is equally straightforward, but the
equations become much more complicated.  We shall present only the case with the two vectors being equal, namely the Lagrangian (\ref{twoequallag}).  The equations are then given again by (\ref{singleom1})-(\ref{singleom2}), but with $a$ replaced by $\tilde a$ and the scalar potential given instead by (\ref{twoequalV}).  We find that the scaling exponent $z$ in gauged supergravities is
\begin{eqnarray}
D=4:&&z=\hbox{real root of $3z^3 + z^2 + 4z-2$} \sim 0.4076,\cr
D=5:&& z=2^{\fft23}-1\sim 0.5874,\cr
D=6:&& z\sim 0.3120,\qquad {\rm or}\qquad z\sim 0.6792,\cr
D=7:&&z\sim 0.4663,\qquad {\rm or} \qquad z\sim 0.7352.
\end{eqnarray}
In $D=6,7$ or higher, there are two solutions for the scaling exponent. All the numbers are less than 1, indicating that it is one of the spatial direction that is singled out in the anisotropic scaling.

For the three and four vector theories in five and four dimensions, the equations of the general Lifshitz ansatz are complicated.  We shall consider only the equal-charge case, for which the theories were generalized to arbitrary dimensions (\ref{einmaxscalar}), and we find
\begin{eqnarray}
ds^2&=& \ell^2 \Big( r^{2z} dy^2 + \fft{dr^2}{r^2} + r^2 dx^\mu dx^\nu \eta_{\mu\nu}\Big)\,,\cr
A &=&\sqrt{\ft{2N}{D-3}} \ell r^z dy\,,\qquad
\cosh^2\fft{\varphi}{\sqrt{N}} = \fft{(D^2-5D+7)(D^2-2D-1)}{(D-3)(D^3-5D^2+7D-1)}\,,\cr
z&=& \fft{(D-3)^2}{D^2-4D+5}\,,\qquad
\ell^2 g^2 = \fft{(D-3)^3(D^3-5D^2+7D-1)}{4(D^2-4D+5)^2}\,.
\end{eqnarray}
Owing to the fact that $z<1$ with $z\rightarrow 1$ as $D\rightarrow \infty$, the solution has anisotropic scaling along a spatial rather than the temporal coordinate.

\subsection{Schr\"odinger vacua}

   For the Schr\"odinger ansatz, we find solutions only for the
single-vector Lagrangian (\ref{singleVlag}) in general dimensions.  There are two branches of solution.  The first branch has $z> 0$, given by
\begin{equation}
ds^2=\ell^2(r^{2z} dt^2 + \fft{dr^2}{r^2} + r^2 (-2dx dt + dy^2))\,,\qquad
A=q r^z dt\,,
\end{equation}
with
\begin{eqnarray}
z&=& \fft{-(D+1)(D-3) + \sqrt{(D+1)(D^3-5D^2+19D-7)}}{2(D+1)}\,,\cr
g^2\ell^2 &=& \ft14(D-1)(D-3)(\fft{D-3}{D+1})^{\fft{D-1}{2(D-2)}}\,,\cr
g^2 q^2 &=& \fft{(D-1)(D-3)\Big(D^2-1 - \sqrt{(D+1)(D^3-5D^2+19D-7)}\Big)}{
2\Big(-(D+1)(D-3) +  \sqrt{(D+1)(D^3-5D^2+19D-7)}\Big)}\,,\cr
e^{\phi} &=& (\fft{D-3}{D+1})^{\fft{D-1}{2(D-2)}}\,,\qquad
\cosh\varphi = \fft{D-1}{\sqrt{(D-3)(D+1)}} >1\,.
\end{eqnarray}
For $D=4$, we have $z>1$ and $q^2<0$, and hence the solution should be analytically continued to become the standard type with $g_{tt}<0$.  For $D\ge 5$, we have $0< z <1$ and $q^2>0$. The scaling exponent $z$ approaches positively to zero as $D\rightarrow \infty$.

The second branch has $z<-(D-3)$, given by
\begin{equation}
ds^2=\ell^2(-r^{2z} dt^2 + \fft{dr^2}{r^2} + r^2 (-2dx dt + dy^2))\,,\qquad
A=q r^z dt\,,
\end{equation}
with
\begin{eqnarray}
z&=& -\fft{(D+1)(D-3) + \sqrt{(D+1)(D^3-5D^2+19D-7)}}{2(D+1)}\,,\cr
g^2\ell^2 &=& \ft14(D-1)(D-3)(\fft{D-3}{D+1})^{\fft{D-1}{2(D-2)}}\,,\cr
g^2 q^2 &=& \fft{(D-1)(D-3)\Big(D^2-1 + \sqrt{(D+1)(D^3-5D^2+19D-7)}\Big)}{
2\Big((D+1)(D-3) +  \sqrt{(D+1)(D^3-5D^2+19D-7)}\Big)}\,,\cr
e^{\phi} &=& (\fft{D-3}{D+1})^{\fft{D-1}{2(D-2)}}\,,\qquad
\cosh\varphi = \fft{D-1}{\sqrt{(D-3)(D+1)}}\,.
\end{eqnarray}
In this case, the scaling exponent $z$ approaches $-\infty$ as $D\rightarrow \infty$. Both the Schr\"odinger vacua are consistent with those of massive vector discussed in section 3.

It is easy to understand the absence of the Schr\"odinger solutions in the two equal-vector theory (\ref{twoequallag}) and the Einstein-Maxwell theory with a hyperscalar (\ref{einmaxscalar}).  The Schr\"odinger ansatz implies that the quantity $A_\mu A^\mu$ vanishes.  It follows that the scalar $\varphi$ is determined by the scalar potential only.  Thus for (\ref{twoequallag}) and (\ref{einmaxscalar}) we must have $\varphi=0$ by the $\varphi$ equation, in which case $A$ becomes massless.

\section{Supergravity-inspired Theories}

In the previous sections, we obtained the truncated Lagrangian in gauged maximal supergravities for the gauge fields that belonged to the Cartan subgroups. We then obtained a class of Lifshitz and Schr\"odinger solutions.  In these solutions, the scaling exponent $z$ is completely fixed in a given solution.  Furthermore, aside from a few examples of rational $z$, most of the solutions has irrational numbers for $z$.  In this section, we shall tweak the supergravity Lagrangians in a mild way so that the theories can admit a more general scaling exponent.  Let us consider the Lifshitz vacua first.  The simplest class of theories in this paper that give rise to Lifshitz solutions is clearly (\ref{einmaxscalar}).  Let us tweak the theory by altering the coupling in the $A^2$ term from $g$ to $\tilde g$, so now the theory consists of two independent couplings $g$ and $\tilde g$.  In other words, we consider the Lagrangian
\be
{\cal L}_D=\sqrt{-g}\Big( R - \ft12(\partial\varphi)^2 - \ft14 F^2 -2 \tilde g^2 \sinh^2(\fft{\varphi}{\sqrt{N}})A^2 + \fft{2Ng^2}{D-3} (D-2 + \cosh\fft{2\varphi}{\sqrt{N}})\Big)\,.\label{superginspire1}
\ee
We can now obtain Lifshitz solutions with the general continuous scaling exponent $z$ instead of it being fixed:
\begin{eqnarray}
ds^2&=& \ell^2 \Big( r^{2z} dy^2 + \fft{dr^2}{r^2} + r^2 dx^\mu dx^\nu \eta_{\mu\nu}\Big)\,,\cr
A &=&\sqrt{\ft{2(1-z)}{z}}\, \ell r^z dy\,,\qquad
\cosh\fft{2\varphi}{\sqrt{N}} = 1 + \fft{(D-1)(D-2)(1-z)}{(z+D-2)(z+D-3)} \,,\cr
\tilde g^2 &=& \fft{2(D-2)z}{(D-3)^2(1-z)}g^2\,,\qquad
\ell^2 g^2 = \fft{(D-3)^2(z+D-2)(z+D-3)}{4(D-1)(D-2)}\,,\label{zle1}
\end{eqnarray}
If we let $\tilde g=g$, we recover the previous result. The reality of the solution requires that $0<z<1$.

    For the application in the condensed matter system, it is perhaps
more interesting to consider the cases with $z> 1$, with the $z=2$ particularly useful.  Although we have constructed Lifshitz solutions with $z>1$ in section 4.1, but they all have $z<2$.  To obtain the $z=2$ solutions, we can tweak the superpotential (\ref{superpot}) as follows
\be
W=\sqrt2 N g \cosh\fft{\alpha\varphi}{\sqrt{N}}\,.\label{superpot1}
\ee
In maximal supergravities in $D=4$ and 5, the constant $\alpha$ is fixed to be unit.  Instead, we can choose a different $\alpha$, solutions with $z=2$ can emerge.  As a concrete example, let us consider $\alpha=2$.  The scalar potential is then given by
\be
V=-\fft{Ng^2}{D-3}\Big(5D-13 - (3D-11)\cosh(\fft{4\varphi}{\sqrt{N}})\Big)\,.
\ee
The Lagrangian is
\be
{\cal L}_D=\sqrt{-g}\Big( R - \ft12(\partial\varphi)^2 - \ft14 F^2 -2 \tilde g^2 \sinh^2(\fft{\varphi}{\sqrt{N}})A^2 -V)\,.
\ee
For the Lifshitz ansatz of the type (\ref{lifsans}), we find $\varphi$ must satisfy
\bea
&&\cosh^2\fft{2\varphi}{\sqrt{N}} - \fft{z^2 +(D-3)z) +2( (D-2)^2}{
(D-2)(2D-5) + 2z^2 + (3D-8)z} \cosh\fft{2\varphi}{\sqrt{N}}\cr
&& -\fft{4(D-3)(D-2)(z-1)}{(3D-11)(2z^2 + (3D-8)z +(D-2)(2D-5))}=0\,.
\eea
It is easy to verify that for real $\varphi$, we must have $z>1$, opposite to the case (\ref{zle1}), corresponding to having $\alpha=1$.  Let us present some explicit low-lying examples of $z=2$:  they all have $q=\ell$ and
\bea
D=4:&& \left\{\ft{1}{96}(39 + \sqrt{69}), \ft{16}{11}(5 + \sqrt{69}), \ft{1}{11}(5 + \sqrt{69})\right\}\,,\cr
D=5:&& \left\{ \ft1{48} (57 + \sqrt{511}), \ft{24}{37} (17 + \sqrt{511}), \ft1{37} (17 + \sqrt{511})\right\}\,,\cr
D=6:&&\left\{ \ft{21}{320} (35 + \sqrt{265}), \ft{8}{9} (13 + \sqrt(265)), \ft{1}{28} (13 + \sqrt{265})\right\}\,,\cr
D=7:&& \left\{\ft{1}{120} (447 + 5 \sqrt{2001}), \ft{25}{79} (37 + \sqrt{2001}), \ft{1}{79} (37 + \sqrt{2001})\right\}\,,
\eea
where the quantities in the lists are those of $\{g^2\ell^2, \tilde g^2/g^2, \cosh\fft{2\varphi}{\sqrt{N}}\}$.

  Another relative simple example that we can tweak is the single-vector
theory given in section 2.2.  We replace the coupling constant $g$ in the $A^2$ term with $\tilde g$, namely
\begin{equation}
e^{-1}{\cal L} = R - \ft12(\partial\phi)^2 - \ft12 (\partial\varphi)^2 -
\ft14 e^{a \phi} F^2 - 2 \tilde g^2 \sinh^2\varphi\, A^2 - V\,,\label{superginspire2}
\end{equation}
with the scalar potential $V$ being the same as (\ref{singleAV}).
we can obtain a general class of Lifshitz solutions with $z$ as a free parameter.  However as in the case of (\ref{zle1}) which requires that $z<1$, the generalized solutions require $z<2$.  We expect that the $z=2$ solution can emerge if we change the structure of the superpotential.

Now let us consider the Schr\"odinger solutions in these supergravity-inspired theories.  Since $A_\mu A^\mu$ vanishes in Schr\"odinger ansatz, the $\varphi$ vanishes in the theory such as (\ref{superginspire1}).  It follows that the Lagrangian (\ref{superginspire1}), even with the superpotential replaced by a more general one (\ref{superpot1}), does not admit Schr\"odinger solutions.  On the other hand, the Lagrangian (\ref{superginspire2}) can admit the Schr\"odinger solution (\ref{schrans1}) with a generic value of $z$ that depends on the ratio $\tilde g^2/g^2$:
\bea
g^2\ell^2 &=& \ft14(D-3)(D-1) e^{a\phi}\,,\qquad
g^2 q^2 = \ft{(D-3)(D^2-1) (z-1)(2z + D-3)}{2(D+1)z^2 + 4(D-1)}\,,\cr
e^{a\phi} &=& \left(\ft{D+1}{D-3}\right)^{-\fft{D-1}{2(D-2)}}\,,\qquad
\cosh\varphi=\ft{D-1}{\sqrt{(D+1)(D-3)}}\,,\cr
\fft{\tilde g^2}{g^2} &=& \fft{(D+1)z(z+D-3)}{4(D-1)}\,.
\eea
The generalized Lagrangian (\ref{superginspire2}) has two non-trivial coupling parameters $(g,\tilde g)$.  The last equation above shows that $z$ is determined by the ratio of these two parameters.  For $z=2$, we have
\be
\fft{\tilde g^2}{g^2} = \ft12 (D+1)\,.
\ee
Thus we see that both Lifshitz and Schr\"odinger solutions with $z=2$ can emerge if we tweak the Lagrangians.  Although these supergravity-inspired theories no longer have the string or M-theory origin,
they are simple generalizations to either the Kaluza-Klein or Einstein-Maxwell theories, and can be used as gravity models to study some non-relativistic condensed system,

\section{Bubbling AdS Solitons}

We have so far considered Lifshitz and Schr\"odinger solutions in the
STU modes in gauged supergravities with additional hyperscalars and their higher-dimensional generalizations.   We now turn to a different application of these theories.

\subsection{Charged AdS black holes and superstars}

The STU models in gauged supergravities also admit another important class of solutions, the static charged AdS black holes \cite{Behrndt:1998jd,Duff:1999gh,tenauthor, Cvetic:1999un, Klemm:2012yg, Klemm:2012vm}.  In the BPS limit, when the mass is a linear summation of charges, the solutions cease to have a horizon, but develop a naked singularity.  To illustrate this, let us consider the Reissner-Nordstr\o m (RN) AdS black hole as an example.  In the $p$-brane coordinates, the solution is given by
\begin{eqnarray}
ds^2 &=& -H^{-2} f dt^2 + H^{\fft2{D-3}} \Big(\fft{dr^2}{f} + r^2 d\Omega_{D-2}^2\Big)\,,\qquad A=\sqrt{N}\,\coth\delta\, H^{-1}  dt\,,\cr
f&=&1 -\fft{\mu}{r^{D-3}} + \ft{4}{(D-3)^2} g^2 r^2 H^N\,,\qquad
H=1 + \fft{\mu \sinh^2\delta}{r^{D-3}}\,,
\end{eqnarray}
where $N$ is given in (\ref{Nvalue}).  The ``BPS'' limit, for which the mass becomes equal to charge (up to a numerical constant depending on the convention,) is given by setting $\mu\rightarrow 0$, $\delta\rightarrow \infty$ while keeping $Q\equiv \mu \sinh^2\delta$ fixed as a finite and non-vanishing quantity. The solution is then given by
\begin{equation}
f=1 + \ft{4}{(D-3)^2} g^2 r^2 H^N\,,\qquad
H=1 + \fft{Q}{r^{D-3}}\,.\label{rnsup}
\end{equation}
It is easy to verify that the space at $r=0$, which for $g=0$ would be the horizon with zero temperature, is not a horizon and the metric has a naked curvature singularity when $H=0$.  This behavior is rather generic for the charged AdS black holes in the STU models and these BPS solutions are called superstars. It turns out that in the single-charge case, the singularity can be resolved via bubbling AdS technique in $D=11$ or $D=10$ \cite{Lin:2004nb}.  For some special AdS bubble configurations, the higher-dimensional solutions can be reduced to become smooth solitons of relevant gauged supergravities, in which the single-vector theory with an additional hyperscalar can be embedded.  In \cite{Chong:2004ce}, multi-charge superstars were resolved by adding multiple hyperscalars.

    As we have discussed in section 2, we obtain the natural
generalizations of the STU modes with hyperscalars in arbitrary dimensions.  These theories without hyperscalars were shown to admit charged AdS black holes \cite{Chow:2011fh,Lu:2013eoa}.  The superstar limit was obtained and discussed in \cite{Lu:2013eoa}.  The question was raised of whether the singularity of such superstars in general dimensions can also be resolved.  In the following subsections, we demonstrate that this can indeed be done with the theories we construct.  In particular, we obtain the smooth soliton that is the resolution of the RN superstar.

\subsection{Single-charge soliton}

The theory is given in section 2.2. The theory with $\varphi=0$ was constructed in \cite{Wu:2011zzh,Liu:2012jra}.  It admits charged AdS black holes that has the superstar limit \cite{Lu:2013eoa}.  With the $\varphi$ turned on, we find
\begin{eqnarray}
ds^2 &=& -H^{-\fft{D-3}{D-2}} f dt^2 + H^{\fft1{D-2}} \Big(\fft{dr^2}{f} + r^2 d\Omega_{D-2}^2\Big)\,,\qquad A=H^{-1} dt\,,\cr
\phi &=& \ft12 a \log H\,,\qquad f=1 + \ft{4}{(D-3)^2}g^2 r^2 H\,,\qquad
\cosh\varphi = (\rho H)'\,,
\end{eqnarray}
where $\rho=r^{D-3}$ and a prime denotes a derivative with respect to $\rho$.   The function $H$ satisfies
\begin{equation}
\rho^{\fft{D-5}{D-3}} (\rho H)'' f + \ft{4g^2}{(D-3)^2} [ (\rho H)'^2 - 1]=0\,.\label{singleHeom}
\end{equation}
The solution can be obtained as follows.  Comparing the equation for $A$ and $\phi$ and then getting rid of the $H''(r)$ term, we find
\begin{equation}
\cosh\varphi = H + \ft{r}{D-3} H_{,r}=(\rho H)'\,.
\end{equation}
The equation (\ref{singleHeom}) then follows straightforwardly from the equation of motion for $A$.  The solutions for $D=4,5,7$ were obtained in \cite{Lin:2004nb}.

 In $D=5$, the equation (\ref{singleHeom}) can be solved exactly \cite{Lin:2004nb}, given by
\be
H=\sqrt{1 + \fft{2m}{\rho} + \fft{q^2}{\rho^2}} - \fft{1}{g^2 \rho}\,.
\ee
When the parameter $m=q$, then $H=1 + (q - g^{-2})/\rho$, and the solution becomes the usual superstar with $\varphi=0$.  For the solution to become a smooth soliton, $H$ must be regular at $\rho=0$, implying that $q=1/g^2$, leading to
\be
H=\sqrt{1 + \fft{2m}{\rho} + \fft{1}{g^4 \rho^2}} - \fft{1}{g^2\rho}\,.
\ee
We must have $m\ge 1/g^2\equiv q$ so that $H$ stays positive definite for $\rho$ running from 0 to $\infty$.

In general dimensions, we have not found any exact solutions. The numerical analysis indicates that the solutions describe smooth solitons with the similar patten of the $D=5$ example.  There exists a parameter choice such that the function $H$ at $\rho=0$ is finite and non-zero with a regular Taylor series expansion for small $\rho$.  At large $\rho$, the function $H$ approaches constant 1 with the two sub-leading terms always being $1/\rho$ and $1/\rho^2$.  In particular, the linearized equation for $u$ defined by $H=1 + \epsilon u$ with $\epsilon\rightarrow 0$ can be solved exactly:
\be
u=\fft{c_1}{\rho} + \fft{c_2}{\rho^2}\, {}_2F_1\Big(\ft12(D-3),(D-3);\ft12(D-1);-\ft{(D-3)^2}{4g^2}
\rho^{-\fft{2}{D-3}}\Big)\,.
\ee

\subsection{Two-charge soliton}

The theory is given in section 2.3.  The charged AdS black holes with vanishing $\varphi_i$ were obtained in \cite{Chow:2011fh} and the superstar limit was discussed in  \cite{Lu:2013eoa}.  Turning on $\varphi_i$, we find
\begin{eqnarray}
ds^2 &=& -(H_1H_2)^{-\fft{D-3}{D-2}} f dt^2 + (H_1 H_2)^{\fft1{D-2}} \Big(\fft{dr^2}{f} + r^2 d\Omega_{D-2}^2\Big)\,,\qquad A_i=H_i^{-1} dt\,,\cr
f&=&1 + \ft{4}{(D-3)^2}g^2 r^2 H_1H_2\,,\quad
\cosh\varphi_i = (\rho H_i)'\,,\quad X_i = (H_1 H_2)^{\fft{D-3}{2(D-2)}} H_i^{-1}
\end{eqnarray}
where $\rho=r^{D-3}$ and a prime denotes a derivative with respect to $\rho$.   The function $H$ satisfies
\begin{equation}
\rho^{\fft{D-5}{D-3}} (\rho H_i)'' f + \ft{4g^2}{(D-3)^2} [ (\rho H_i)'^2 - 1]H_1 H_2 H_i^{-1}=0\,.\label{twoHeom}
\end{equation}
Note that the solutions for $X_i$ can be equivalently expressed as
\begin{equation}
\vec \phi= \ft12 \vec a_1 \log H_1 - \ft12 \vec a_2 \log H_2\,.
\end{equation}
The solutions with $D=4,5,6,7$ were obtained in \cite{Chong:2004ce}.  Numerical analysis indicate that the solutions describe smooth solitons whose asymptotic infinities match those of the corresponding superstars.  The two sub-leading orders are of the structure $1/\rho$ and $1/\rho^2$. In particular, the linearized equation of $H_i$ around 1 has the same equation as the one in single charge case, discussed previously.

\subsection{Soliton resolutoin of the RN superstar}

The theory is given by (\ref{einmaxscalar}). For vanishing $\varphi$, it becomes Einstein-Maxwell gravity with a cosmological constant.  The RN AdS black hole and its superstar limit were discussed in section 6.1.  We find
\begin{eqnarray}
ds^2 &=& -H^{-2} f dt^2 + H^{\fft2{D-3}} \Big(\fft{dr^2}{f} + r^2 d\Omega_{D-2}^2\Big)\,,\qquad A=\sqrt{N}\,H^{-1} dt\,,\cr
f&=&1 + \ft{4}{(D-3)^2}g^2 r^2 H^N\,,\quad
\cosh\fft{\varphi}{\sqrt{N}} = (\rho H)'\,,
\end{eqnarray}
where $\rho=r^{D-3}$ and a prime denotes a derivative with respect to $\rho$.   The function $H$ satisfies
\begin{equation}
\rho^{\fft{D-5}{D-3}} (\rho H)'' f + \ft{4g^2}{(D-3)^2} [ (\rho H)'^2 - 1]H^{N-1}=0\,.\label{equalHeom}
\end{equation}
It is clear that if we let $\varphi=0$, which requires that $H=1 + Q/\rho$, the solution becomes the RN superstar (\ref{rnsup}).  In general, the solution describe a smooth soliton with $H$ runs from a constant at $\rho=0$ to $1+c_1/\rho + c_2/\rho^2 + \cdots$ at the asymptotic AdS infinity.  This result implies that the bubbling fermi droplets in super conformal Yang-Mills theories in four dimensions may exist in general dimensions.

\section{Conclusions}

In this paper, we studied gauged maximal supergravities via the sphere reduction of eleven-dimensional and type IIB supergravities.  We considered the subset of fields, namely the full $SO(n+1)$ vectors $A^{ij}$ and the scalars $T^{ij}$ that parameterize the sphere deformation.  We note that sphere reduction ansatz takes a universal form for this subset.  The consistency of the reduction requires a further constraint (\ref{ffcons}) on the gauge fields, which is satisfied for the solutions we constructed in this paper.

    We then focused on a further consistent truncation to the abelian gauge
fields associated with the Cartan subgroup and the diagonal entries of the scalars $T^{ij}$.  These theories describe the STU modes in gauged supergravities with additional hyperscalars \cite{Chong:2004ce}. With only one or two $U(1)$ fields turned on, we find that the theory can be naturally generalized to arbitrary dimensions.  In four and five dimensions, when all the $U(1)^4$ or $U(1)^3$ gauge fields are set to equal, the theory reduces to Einstein-Maxwell gravity with one hyperscalar.  We find that it can also be generalized naturally to arbitrary dimensions.

   The hyperscalars turn out to provide mass terms for the gauge fields
analogous to the Abelian Higgs mechanism. In the supersymmetric AdS vacua, the hyperscalars vanish and the gauge fields are massless.  We constructed a class of Lifshitz and Schr\"odinger vacua in which the hyperscalars are non-vanishing and play an important role.  However, gauged maximal supergravities have only one non-trivial parameter $g$ and the scalar potential associated with diagonal entries of the $T^{ij}$ has few extrema. The scaling exponent $z$ in our solutions is severely restricted, taking fractional or even irrational numbers.  We tweaked the theories by introducing a new coupling constant $\tilde g$ to the mass term and/or changing the scalar superpotential slightly.  We then obtain Lifshitz and Schr\"odinger solutions with a generic exponent, including $z=2$.  The tweaked theories are simple generalizations of
the Einstein-Maxwell or Kaluza-Klein theories and can be used to study some non-relativistic condensed matter system.

   In a different application, we use our new theories with hyperscalars
to obtain smooth solitons that are resolutions of singular superstars that arise commonly as the BPS limits of the charged AdS black holes in STU models.  In particular, we obtain the soliton resolution of the RN superstars in general dimensions.  The existence of such solution justifies the ``naturalness'' of our generalizations of the supergravity Lagrangians to higher dimensions.  It may indicate that the picture of the AdS bubbling fermion droplets found in the four-dimensional superconformal Yang-Milll theories may exist in higher dimensions.

    To summarize, we obtain a class of Lifshitz and Schr\"odinger vacua
in maximal gauged supergravities. We also obtain the higher-dimensional generalizations of these theories that admit such vacua.  It turns out that these theories also admit smooth bubbling AdS solitons that have played some important role in the AdS/CFT correspondence.

\section*{Acknowledgement}

H-S.L.~is supported in part by the NSFC grant 11305140 and SFZJED grant Y201329687. H.L.~is supported in part by the NSFC grants 11175269 and 11235003.

\end{document}